\documentclass[12pt]{article}
\usepackage{amsmath}
\usepackage{amsfonts}
\usepackage{graphicx,psfrag,epsf}
\usepackage{enumerate}
\usepackage{natbib}
\usepackage{float}
\usepackage{url} 
\usepackage{setspace}
\usepackage{algorithm}
\usepackage{algpseudocode}
\usepackage{hyperref}
\usepackage[mathscr]{euscript}
\usepackage{xcolor}
\usepackage{authblk}

\newcommand{\blind}{1}

\begin{document}

\def\spacingset#1{\renewcommand{\baselinestretch}%
{#1}\small\normalsize} \spacingset{1}
\newcommand{\jsm}[1]{\textcolor{blue}{\textsf{#1}}}
\newcommand{\cut}[1]{\textcolor{red}{\textsf{#1}}}


\if1\blind
{
\title{\LARGE\bf Semiparametric quantile functional regression analysis of adolescent physical activity distributions in the presence of missing data}
\author[1,2]{Benny Ren}
\author[2]{Ian Barnett}
\author[2]{Haochang Shou}
\author[2]{Jeremy Rubin}
\author[4]{Hongxiao Zhu}
\author[3]{Terry Conway}
\author[3]{Kelli Cain}
\author[5]{Brian Saelens}
\author[2]{Karen Glanz}
\author[3]{James Sallis}
\author[2]{Jeffrey S. Morris}
\affil[1]{Department of Family, Population and Preventive Medicine, Renaissance School of Medicine at Stony Brook University}
\affil[2]{Department of Biostatistics, Epidemiology and Informatics, 
Perelman School of Medicine at the University of Pennsylvania}
\affil[3]{Herbert Wertheim School of Public Health and Human Longevity Science, University of California San Diego}
\affil[4]{Department of Statistics, Virginia Tech}
\affil[5]{Department of Pediatrics, University of Washington School of Medicine}
\maketitle
} \fi

\if0\blind
{
  \bigskip
  \bigskip
  \bigskip
  \begin{center}
    {\LARGE\bf Semiparametric quantile functional regression analysis of adolescent physical activity distributions in the presence of missing data}
\end{center}
  \medskip
} \fi

\bigskip
\begin{abstract}
In the age of digital healthcare, passively collected physical activity profiles from wearable sensors are a preeminent tool for evaluating health outcomes. In order to fully leverage the vast amounts of data collected through wearable accelerometers, we propose to use quantile functional regression to model activity profiles as distributional outcomes through quantile responses, which can be used to evaluate activity level differences across covariates based on any desired distributional summary. {Our proposed framework addresses two key problems not handled in existing distributional regression literature.  First, we use spline mixed model formulations in the basis space to model nonparametric effects of continuous predictors on the distributional response.  Second, we address the underlying missingness problem that is common in these types of wearable data but typically not addressed.  We show that the missingness can induce bias in the subject-specific distributional summaries that leads to biased distributional regression estimates and even bias the frequently used scalar summary measures, and introduce a nonparametric function-on-function modeling approach that adjusts for each subject's missingness profile to address this problem.}
{We evaluate our nonparametric modeling and missing data adjustment using simulation studies based on realistically simulated activity profiles and use it to gain insights into adolescent activity profiles from the Teen Environment and Neighborhood study.}
\end{abstract}

\noindent%
\vfill

\newpage
\spacingset{1.45} 
\section{Introduction}
\label{sec:intro}

Physical activity is a well-documented protective factor for prevention and treatment of numerous physical and mental health conditions \citep{rowlands2019enhancing,bell2019relationship} Accelerometers provide validated objective measures of physical activity and are widely used in research \citep{migueles2022granada}. However, accelerometers generate streams of data that are complex, and there is a substantial methodological literature on developing optimal approaches to collecting and scoring accelerometer data \citep{van2022associations,catellier2005imputation}. For example, accelerometer devices have been used to study the relationship between demographic variables and physical activity in patients \citep{arvidsson2019measurement}. Though longitudinally measured accelerometer counts are naturally modeled as stochastic processes \cite{morris2006using}, they are also often framed as time-invariant distributions whose summary statistics are used to study health outcomes \citep{cain2013using}.

The Teen Environment and Neighborhood (TEAN) study of built environments and physical activity was conducted to study the social-economic and demographic determinants of physical activity \citep{sallis2018neighborhood,borner2018latent,thornton2017relation,carlson2017within}. The TEAN study recorded aggregated accelerometer count data over several days to study a wide range of correlates of adolescent physical activity, including demographic, psychological, social, and neighborhood environment. Typically, this collection of accelerometry data is reduced to scalar summary statistic outcomes related to physical activity and are analyzed under independent models \citep{cain2013using}. For example, measurements such as minutes of \textit{moderate-to-vigorous physical activity} (MVPA) per day have been used to quantify physical activity \citep{evenson2008calibration}. In general, daily summary statistics do not fully leverage complex temporal dependence and diurnal trends that require a more general framework for inference of activity profiles. 

A flexible and powerful alternative to summary statistics involves modeling count data as distributions over the day \citep{ghosal2023distributional,zhang2022two} which are then considered as functional responses. Recent developments in statistical Wasserstein methods have resulted in approaches involving minimizing risk based on quantile functions \citep{chen2021wasserstein,panaretos2019statistical}. Convergence of the quadratic Wasserstein distance is equivalent to convergence in law of two measures \citep{panaretos2019statistical} and motivates the use of a L2 loss function for constructing estimation equations. In addition, the quadratic Wasserstein loss can be decomposed through truncated Karhunen-Loeve (KL) expansions allowing for functional regression to estimate distributional outcomes as quantile functions \citep{yang2020quantile,yang2020random}. Given quantile function estimates, any \textit{functionals} of a distribution \citep{fisher2021visually} can be computed to yield distributional moments, MVPA statistics or any other desired distributional summary statistics.  A primary goal of this paper is to build functional response regression models for subject-specific activity distributions to characterize how activity levels vary across covariates, while accounting for a key missing data problem common in these types of data but unaddressed in current literature. This enables us to evaluate the complex interplay of of sex, BMI, and age on the degree and diurnal nature of physical activity in a sample of adolescent participants.

We use mixed effect components within the functional response regression framework to estimate nonparametric smoothing splines, or generalized additive models (GAMs) \citep{hastie1987generalized}, through hierarchical random effects \citep{wand2008semiparametric,lee2018bayesian} to allow nonlinear covariates effects on the distributional responses.  This construction allows for nonlinear effects of continuous predictors such as BMI to change local regions of the quantile function resulting in inference involving a smooth quantile-by-BMI surface. This is an important improvement over functional coefficients \citep{ghosal2023distributional,yang2020quantile} by allowing standard errors to capture the heterogeneity of the empirical data. For example, there few subjects in the study cohort with a BMI below the 10th percentile and equivariance assumptions across BMI are unreasonable. After model fitting, we can map the quantile-by-BMI surface to summary statistics through functionals of the quantile at conditional values of BMI to yield nonparametric BMI effects on any desired distributional summary, which inherit the smoothness properties of the quantile-by-BMI surface.  Thus, our modeling framework provides a unified approach for semiparametric distributional regression that precludes the need to fit separate GAMs to different distributional summaries.

In addition, these activity data, like many other wearable sensor data, are characterized by intermittent segments of missing data in the time-of-day activity profiles. As we will demonstrate, failure to account for these missing data patterns can lead to biased accelerometer outcomes \citep{cain2013using}. Each subject has a routine missing data pattern coinciding with the removal of accelerometer devices due to factors such as mandated sporting regulations \citep{catellier2005imputation} and bedtimes or when battery charge runs out. Though imputation can be used to recover missing accelerometer data when missing at random (MAR) \citep{catellier2005imputation,morris2006using,yue2018statistical}, these assumptions may not always apply, and certain missingness patterns may affect the subject-specific activity distribution in different ways.  For example, accelerometer counts near bed-time may tend to correspond to lower values in the left tail of the distribution, causing missingness during this time to decrease the amount of sedentary time in the subject-specific distribution and shifting it to the right. In order to account for missing data patterns and study their effect on activity distributions, we propose a joint model of subject-specific activity distribution and missingness profile over time that induces a functional covariate adjustment procedure effectively normalizing all subject-specific distributions to a common missingness pattern. We deconstruct baseline missing data patterns as functional principal components (fPCs) and adjust for missing data through function-on-function regression \citep{yao2005functionala,yao2005functionalb,morris2015functional}. Similar restrictions and applications of nonparametric functional regression are found in many distributional data settings, e.g. financial data, heart rate, climate, etc. \citep{zhang2022wasserstein,ghosal2023distributional,yang2020random}. 

The remainder of the article is organized as follows. In Section \ref{sec:data} we describe the TEAN study data, and preprocessing procedures such stabilizing transformations and missing pattern covariates. In Section \ref{sec:meth} we outline a Bayesian mixed models for function-on-function regression with nonparametric smoothing splines. In Section \ref{sec:sim}, we present extensive simulations to demonstrate the bias due to missing data on quantile estimation. In Section \ref{sec:real}, we present the TEAN study real data analysis.

\section{Data Preprocessing}
\label{sec:data}

Accelerometer data were collected from the Teen Environment and Neighborhood (TEAN) study of built environments and physical activity conducted in two US geographic sites (Baltimore, MD/Washington, DC and Seattle/King County, WA) during 2009–2011 \citep{sallis2018neighborhood,carlson2017within}. A cohort of $N=750$ adolescents ages 12--17 were asked to wear an Actigraph accelerometer model 7164 (acceleration counts were recorded at 30-second epochs) \citep{kozey2010comparison}. Under the \textit{30 minute rule}, segments of $\geq60$ consecutive 30-second epochs (i.e., 30 minutes or greater) with $\text{Count} = 0$ were considered missing data \citep{cain2013using}. We only consider weekdays with $\geq8$ hours of valid wear time and participants with at least 3 valid weekdays {for inclusion in this analysis}. Considering a conservative interval of wake time for adolescents during weekdays, we focus our analysis on accelerometer data from weekdays, 6am--11:30pm \citep{carskadon2011sleep}. A heat map of accelerometer counts ordered by age and time-of-day is plotted in Figure \ref{fig:heat}. For the study population, we have demographic covariates: sex, site, age and BMI percentile. Our goal in this analysis is to characterize how the subjects' activity level distributions vary according to these covariates. 

\subsection{Calculation of Subject-specific Empirical Quantile Functions}
\begin{figure}[H]
\centering
\singlespacing
\includegraphics[width=5in]{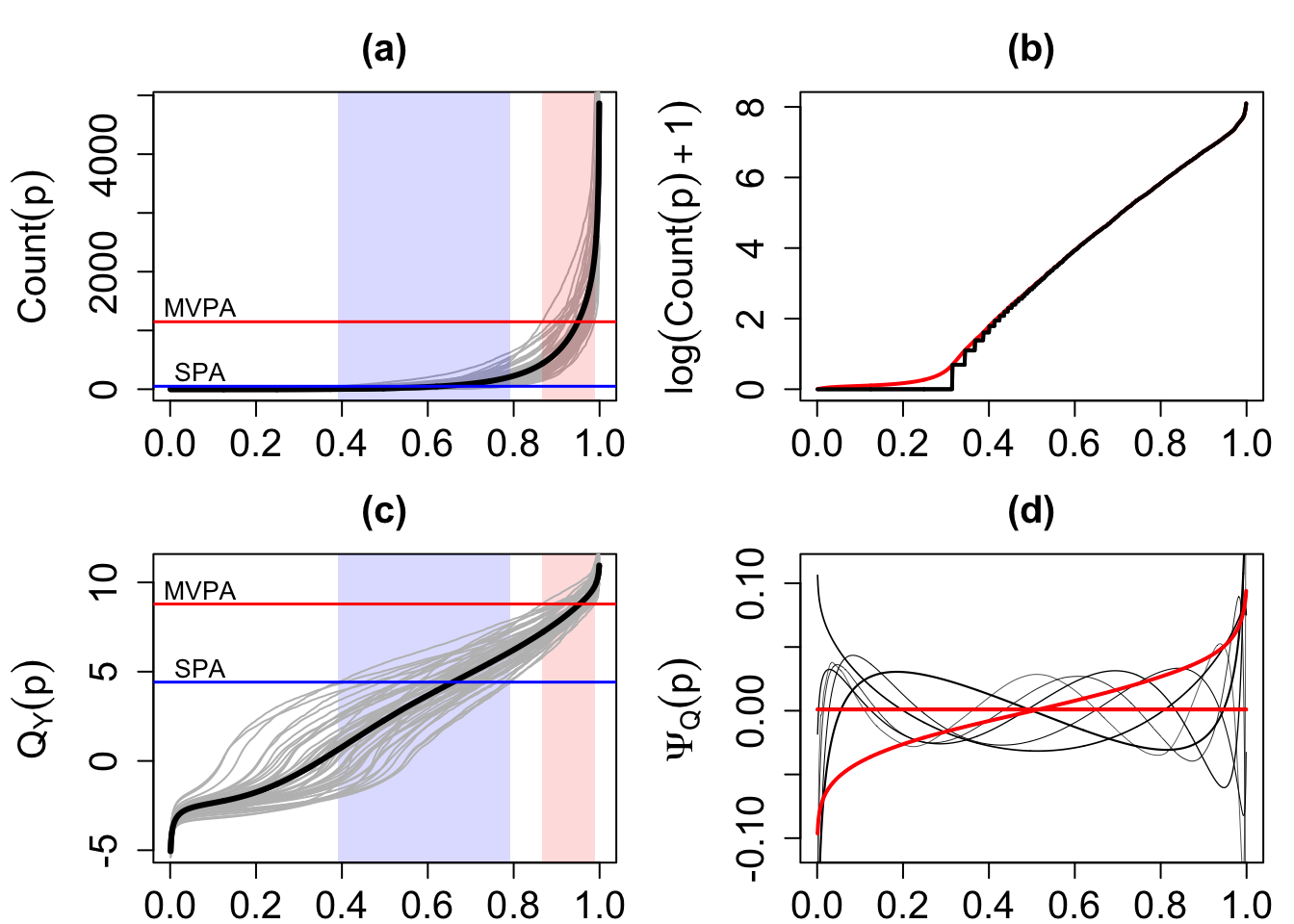}
\caption{Empirical and Box-Cox transformed quantiles. (a) Spaghetti plot of empirical quantiles of count distributions with the Frechet mean plotted in bold black and cutoffs for moderate-vigorous physical activity (MVPA) and sedentary physical activity (SPA). Observed proportions of MVPA and SPA are shaded in red and blue respectively. (b) Example of smooth monotonic interpolation of a discrete value empirical quantile function plotted on the log scale. (c) Spaghetti plot of empirical quantiles of count distributions after smooth Box-Cox transformation with the Frechet mean plotted in bold black and cutoffs for MVPA and SPA. (d) Quantlet orthogonal basis system ($K=8$) used to reconstruct the empirical quantile distributions with the first two quantlets plotted in red.}
\label{fig:BC}
\end{figure}

For each subject $i\in\{1,\dots,N\}$, we have $n_i$ non-negative integer accelerometer count observations: $\mathrm{Count}_i$ recorded at 30-second epochs and can be used to empirically calculate quantile functions and CDFs. The distribution of counts is inherently zero-inflated and right skewed (see Figure \ref{fig:BC}a). In order to facilitate functional data analysis methods, we propose an order preserving \textit{smooth Box-Cox transformation} to stabilize the variance and map from non-negative integers to the real number line, $\mathbb{N}_{\geq 0} \mapsto \mathbb{R}$. 

For subject $i$, the empirical quantile function $\mathrm{Count}_i (p)$ can be calculated as ordered statistics $\mathrm{Count}_{i(1)} \leq \cdots \leq \mathrm{Count}_{i(n_i)}$, with the corresponding domain $p \in \{ 1 /(n_i+1), 2 /(n_i+1), \dots, n_i /(n_i+1) \}$. Our zero-inflated discrete data needs to be smoothed in order to make it amendable to Wasserstein based techniques \citep{petersen2016functional,chen2021wasserstein}. We first compute the \textit{discrete value quantiles} as step functions given by calculating the coupling of function output and domain as $\{ \mathrm{Count}_i(p), p \} = \{ k, \inf\{ p: \mathrm{Count}_i(p) = k \} \}$ which are calculated for each integer $k \in \mathbb{N}_{\geq 0}$. A shape preserving monotonic cubic interpolation of \cite{borchers2023pracma} is next applied to smooth the subject-specific empirical {discrete value quantiles} as $Q_{\mathrm{Count}_i}(p)$ (see Figure \ref{fig:BC}b for an example). This approach preserves the empirical cut-points, and monotonically smooths the in between values, which can be viewed as an interpolation to values we would expect had the measurement instrument had greater precision, accounting for the zero inflation and discreteness at lower counts to yield a continuous-valued monotonic smoothed subject-specific empirical quantile function (Figure \ref{fig:BC}a \& c).
The smoothed quantile functions for all subjects are then mapped to a common uniformly spaced $n=1024$ point common grid of $p \in[1 /(n+1), \dots n /(n+1)]$, denoted as $Q_{\mathrm{Count}_i}(p)$ (see Figure \ref{fig:BC}a), with all values guaranteed to be non-negative, and then a Box-Cox power transform can be applied to make the distributions more regular.  We then propose to model $Q_{\mathrm{Count}_i}(p)$ as a functional response that is regressed on covariates. 

To perform functional regression, we adopt the basis expansion representation of quantile functions \citep{yang2020quantile}. Due to the difficulty of obtaining a KL expansion in Wasserstein space, we propose the use of a Box-Cox transformation \citep{box1964analysis} to stabilize the variance while preserving the order statistics (Figure \ref{fig:BC}a \& c). In order to efficiently train the Box-Cox transformation, we use the Frechet mean of the quantiles to train the Box-Cox transformation. The Frechet mean is given as $\overline{Q}_{\mathrm{Count}}(p) = N^{-1} \sum_{i=1}^N Q_{\mathrm{Count}_i}(p)$ and is a natural center of all empirical quantiles \citep{chen2021wasserstein}. The $n=1024$ values of $\overline{Q}_{\mathrm{Count}}(p)$ are treated as random observations where a Box-Cox transformation can be trained to normalized the Frechet mean. This Box-Cox transformation is then applied to all quantiles as $Q_{Y_i}(p) = \mathrm{BC} \{ Q_{\mathrm{Count}_i}(p) \}$ in order to address right skewed distributions of counts and stabilize the lower moments. In addition, the Box-Cox transformation preserves the registration of $p$ for cut-points and are plotted in Figure \ref{fig:BC}c. Our procedure transforms all quantile functions to be closer to a Gaussian distribution. Notably, this mapping to the real number line is an improvement to the standard log transformation found in the literature \citep{ghosal2023distributional} by removing the positivity constraint of the support (Figure \ref{fig:BC}b). Pseudocode for the \textit{smooth Box-Cox transformation} is presented in the supplement. 

\subsection{Missing Rate Functional Covariates}


The subject-specific distributional summaries are based on all observed (i.e. non-missing) counts, but as mentioned above, there are missing intervals that can potentially bias the subject-specific distributional estimate and any downstream distributional summary statistics. As missing values certain regions of time of day $t$ may be associated with low count values, e.g. \textit{bed-times} in late nights and early mornings, or high count values, e.g. mid-day values, it is important to retain the missingness profile as function of time of day, and adjust for them in the distributional regression analysis. 

Suppose that $R_{ij}(t)$ is a missingness indicator for whether time $t$ for day $j$ for subject $i$ is missing, assuming $R_{ij}(t)\sim$Ber$\{\pi_i(t)\}$, with $\pi_i(t)$ characterizing the subject-level missingness profile.  We will estimate $\pi_i(t)$ for each subject and use it as a functional covariate to calibrate their empirical activity distribution. For interpretability, we will estimate $\pi_i(t)$ on a grid of $t$ involving 35 half-hour intervals from 6am to 11:30pm \citep{cain2013using}. The aggregation of $R_{ij}(t)$ across 60 30-second epochs and at least 3 days of data per subject provides stable estimates of $\pi_i(t)$ on the specified grid. In general, a different aggregation scheme can be used depending on the domain of the modeling problem. We then transform $\pi_i(t)$ onto the real line via $M_i(t) = \operatorname{elogit}\left\{ \pi_{i}(t) \right\}$, with $\operatorname{elogit}(x) = \log \left\{ ({x+ 0.5/N})/({1-x+ 0.5/N}) \right\}$ the empirical logit function \citep{gart1985effect}, to facilitate downstream functional regression modeling.

\section{Methods}
\label{sec:meth}

We use a functional data analysis approach to model $Q_{Y_i}(p)$ and $M_i(t)$, with the observed functions on discrete sampling grids represented as $\boldsymbol{Q}_{i} \in \mathbb{R}^{1024}$ and $\boldsymbol{M}_i \in \mathbb{R}^{35}$, respectively.
The joint distribution $f\{ Q_{Y_i}(p), M_i(t) \mid \boldsymbol{X}_i \}$,
providing functional regression for the subject-specific empirical quantile function and missingness profiles, can be written via the following conditional models:
\begin{equation} \label{eq:factors1}
\begin{array}{rl}
    f\{ Q_{Y_i}(p) \mid \boldsymbol{X}_i, M_i(t) \} =& f ( \boldsymbol{Q}_i \mid \boldsymbol{X}_i, \boldsymbol{M}_i ) \approx f ( \boldsymbol{Q}^*_i \mid \boldsymbol{X}_i, \boldsymbol{M}^*_i ) \\
     f\{ M_i(t) \mid \boldsymbol{X}_i \} =& f ( \boldsymbol{M}_i \mid \boldsymbol{X}_i )
    \approx f ( \boldsymbol{M}^*_i \mid \boldsymbol{X}_i )
\end{array}    
\end{equation}
with coefficients of a quantlet system of basis functions given as $\boldsymbol{Q}^*_i \in \mathbb{R}^{K_Q}$ and coefficients of a functional principal component (PC) system given as $\boldsymbol{M}_i^* \in \mathbb{R}^{K_M}$, following the PACE algorithm \cite{yao2005functionala,yao2005functionalb}. We motivate our semiparametric functional regression approach in the basis space using the $f\{ Q_{Y_i}(p) \mid \boldsymbol{X}_i, M_i(t) \}$ model.

\subsection{Accelerometer Quantile Functional Regression Model}
\label{sec:quantile}

Methods based on the Wasserstein metric have be proposed to construct estimating equations for quantile functional regression  \citep{chen2021wasserstein,zhang2022wasserstein}. These are approaches are well justified by drawing on Wasserstein geometry \citep{panaretos2019statistical}. Convergence in law of two measures is equivalent to the Wasserstein distance converging to zero, providing justification for the quadratic Wasserstein distance to serve as an empirical risk. The approach of \cite{yang2020quantile} based on minimizing the quadratic Wasserstein distance used 
\begin{equation} \label{eq:wd2}
\begin{array}{c}
d_{W}^{2}\left( Q_{Y_i}(p), \sum_{k=1}^K {Q}^*_{i,k} \Psi_{Q,k}(p) \right) = \int_{0}^{1} \left( Q_{Y_i}(p) - \sum_{k=1}^K {Q}^*_{i,k} \Psi_{Q,k}(p) \right)^{2} d p = \| \boldsymbol{Q}_i - \boldsymbol{Q}^*_i \boldsymbol{\Psi}_Q \|^2_2
\end{array}
\end{equation}
to obtain a sparse set of smooth orthogonal basis functions such that ${Q}_{Y_i}(p) = \boldsymbol{Q}_i=\boldsymbol{Q}^*_i \boldsymbol{\Psi}_Q$. Under this approach, we use construct a smooth orthogonal $K_Q=8$ \textit{quantlet} basis expansion of empirical quantiles $Q_{Y_i}(p)$ that has a minimum \textit{leave-one-out concordance correlation coefficient} that is greater than $0.99$ {across \textit{all} $N=750$ subjects}, computed using cross-validation, to ensure a near-lossless KL expansion. Our quantlet basis system is plotted in Figure \ref{fig:BC}d, and described in detail in the supplement. The first two quantlets provide a sufficient basis for all Gaussian quantile functions, with the first a constant function whose coefficient is the mean and the second a standard normal quantile function whose coefficient is the variance for Gaussian data, and the subsequent empirically estimated quantlets capturing non-Gaussian behavior in the empirical quantile functions. 

Similarly, the quadratic Wasserstein distance loss can be used to specify a regression model for $f ( \boldsymbol{Q}_i \mid \boldsymbol{M}_i, \boldsymbol{X}_i )$ using the quadratic risk defined in \eqref{eq:wd2}. The semiparametric function-on-function regression model is given by 
\begin{equation} \label{eq:quantileFMM}
\begin{array}{rl}
Q_{Y_i}(p) =& \boldsymbol{X}_i^{\top} \boldsymbol{\beta}(p) + \int M_i(t) B_M(t,p) dt + S_{\text{age}} \left({\text{age}_i}, {p}\right) + S_{\text{BMI}} \left({\text{BMI}_i}, {p}\right) + E_i(p) \\
\boldsymbol{Q}_{i} =& \boldsymbol{X}^\top_i \boldsymbol{\beta} + \boldsymbol{M}_i \boldsymbol{\beta}_{\text{miss}} + \boldsymbol{Z}_{\text{age},i}^\top \boldsymbol{U}_{\text{age}} + \boldsymbol{Z}_{\text{BMI},i}^\top \boldsymbol{U}_{\text{BMI}} + \boldsymbol{E}_{i}
\end{array}
\end{equation}
where $\boldsymbol{X}_i=[ \boldsymbol{X}^\dagger_i, \text{age}_i, \text{BMI}_i] \in \mathbb{R}^{K_X}$ and $\boldsymbol{X}^\dagger_i$ are the cell mean form of sex-by-site combinations, for a total of $K_X=6$ fixed effects. The nonparametric function $S_{\text{age}} \left(\text{age}_i, {p}\right)$ represents the effect of age on the quantile function output $Q_{Y_i}(p)$ at specific $p$ with the analogous formulation for $S_{\text{BMI}} \left(\text{BMI}_i, {p}\right)$ \citep{lee2018bayesian}. Here $\{\boldsymbol{Z}_{\text{age},i}, \boldsymbol{Z}_{\text{BMI},i} \}$ are Demmler-Reinsch (DR) spline matrices \cite{demmler1975oscillation} derived from second order penalized cubic B-splines and $\{\boldsymbol{U}_{\text{age}}, \boldsymbol{U}_{\text{BMI}} \}$ are random effect coefficients that induce smoothness {across age and BMI}. The nonparametric effect $\int M_i(t) B_M(t,p) dt$ models changes in the $Q_{Y}(p)$ due to missing data pattern $\pi(t)$. Under vector notation, let $\boldsymbol{\beta} = \boldsymbol{\beta}(p) = [\beta_{1}(p),\dots, \beta_{6}(p)]^\top = [\mu_1(p), \mu_2(p), \mu_3(p), \mu_4(p), \beta_{\text{age}}(p), \beta_{\text{BMI}}(p)]^\top \in \mathbb{R}^{K_X \times K_Q}$ is a column vector containing unknown fixed functional coefficients for percentile $p$. The residual error processes $E_i(p)$ are assumed to follow a mean-zero Gaussian process with the covariance surface, $\Sigma(p_1, p_2) = \mathrm{cov}\{E_i(p_1), E_i(p_2)\}$ and to be independent of $\boldsymbol{X}_i^{\top}$ and nonparametric terms. 

We propose covariate adjustment of the missing data patterns following the parameterization of \cite{yao2005functionalb}. We represent $\int M_i(t) B_m(t,p) dt$ by regressing quantlet coefficients $\boldsymbol{Q}^*_i \in \mathbb{R}^{K_Q}$ on $\boldsymbol{M}^*_i \in \mathbb{R}^{K_M}$ where $K_Q=8$ and $K_M=18$ and take advantage of smooth basis systems for functional predictors and outcomes. When performing inference, by setting $\boldsymbol{M}^*=\mathbf{0}$ we condition on the mean empirical missing rate to estimate the quantile function, thus effectively normalizing {each subject's empirical quantile function to a common missing data pattern, which we hypothesize will adjust for potential bias from the missing data}. Using a quantlet basis matrix mapping through $\boldsymbol{\Psi}_Q$, we can then write the $k$th basis coefficient of $\boldsymbol{Q}^*_i$ as regression
\begin{equation} \label{eq:quantilecoef}
\begin{array}{c}
{Q}^*_{i,k} = \boldsymbol{X}^\top_i \boldsymbol{\beta}^*_k + \boldsymbol{M}^*_i \boldsymbol{\beta}^*_{\text{miss},k} + \boldsymbol{Z}_{\text{age}_i}^\top \boldsymbol{U}^*_{\text{age},k} + \boldsymbol{Z}_{\text{BMI}_i}^\top \boldsymbol{U}^*_{\text{BMI},k} + {E}^*_{i,k}
\end{array}
\end{equation}
where $\boldsymbol{U}^*_{\text{age},k} \sim \mathrm{N}(\mathbf{0}, \tau^{\star}_{\text{age},k}\mathbf{I}_{J+2})$, $\boldsymbol{U}^*_{\text{BMI},k} \sim \mathrm{N}(\mathbf{0}, \tau^{\star}_{\text{BMI},k}\mathbf{I}_{J+2})$ {and ${E}^*_{i,k}\sim\mathrm{N}(\mathbf{0},s^*_k)$ are independent normally distributed residual components.  Importantly, note that the heteroscedastic assumptions for $E^*_{i,k}$ induces flexible functional covariance surfaces for the curve-to-curve deviations $\Sigma(p_1,p_2)=\mathrm{cov}\{E_i(p_1),E_i(p_2)\}$ because of the structure of the learned quantlets $\boldsymbol{\Psi_Q}$.} We elaborate how mixed effects $\boldsymbol{U}^*$ can be used to fit a smooth nonparametric effect in the supplement. See \cite{wand2008semiparametric,lee2018bayesian,o1986statistical,speed1991blup} for additional details. Coefficients vectors $\{ \boldsymbol{U}^*_{\text{age},k}, \boldsymbol{U}^*_{\text{BMI},k} \}$ are stacked to form matrices $\{ \boldsymbol{U}^*_{\text{age}}, \boldsymbol{U}^*_{\text{BMI}} \}$ and are mapped back to $\{ \boldsymbol{U}_{\text{age}}, \boldsymbol{U}_{\text{BMI}} \}$ using $\boldsymbol{\Psi}_M$. Our model can also be interpreted as penalizing $\int\left\{S_v^{\prime \prime}(v, {p})\right\}^2 d v $ for our nonparametric functions at values of $v \in \{\text{age},\text{BMI}\}$. The smoothness induced by mixed effect modeling in the coefficient space $\boldsymbol{Q}^*$ results in a smooth nonparametric regression in the original domain $Q_Y(p)$ (see \cite{lee2018bayesian} for an additional example in functional regression). Through second derivative penalties, our model induces smoothness for across nonparametric functions of age and BMI at each principal subspace of $p$ and additionally induces smoothness across $p$ by the local smoothing of the quantlet construction (Figure \ref{fig:BC}d), with the smoothness across age and BMI tending to itself be smooth across $p$. This enables adaptive borrowing of strength across the covariates and $p$ in determining the two-dimensional smooth coefficient. 

The BayesFMM framework first introduced by \cite{morris2006wavelet} can be used to fit \eqref{eq:quantilecoef} and obtain posterior draws $\boldsymbol{Q}^{(r)} \sim f ( \boldsymbol{Q} \mid \widetilde{\boldsymbol{M}}, \widetilde{\boldsymbol{X}} )$. 
Note that we abbreviate $\{ \widetilde{\boldsymbol{X}}, \widetilde{\boldsymbol{Z}}_{\text{age}}, \widetilde{\boldsymbol{Z}}_{\text{BMI}} \}$ as $\widetilde{\boldsymbol{X}}$ because $\{ \widetilde{\boldsymbol{Z}}_{\text{age}}, \widetilde{\boldsymbol{Z}}_{\text{BMI}} \}$ are spline mappings and functions of $\{ \text{age},\text{BMI} \}$ from $\widetilde{\boldsymbol{X}}$. The regularized smooth surface of $f\{ Q_Y(p) \mid \widetilde{\boldsymbol{X}} \}$ mitigates the influence of outliers during estimation. A analogous model for $f\{ M_i(t) \mid \boldsymbol{X}_i \}$ can be fit if the effect of covariates on the missingness function itself is of interest, and is presented in the supplement. Given posterior draws from the quantlet model (\ref{eq:quantilecoef}), we can compute posteriors back in the $p$ domain (\ref{eq:quantileFMM}) 
via ${Q}_{Y_i}^{(r)}(p) = \boldsymbol{Q}^{(r)}_i =\boldsymbol{Q}^{*(r)}_i \boldsymbol{\Psi}_Q$. This estimation procedure also ensures a smooth nonparametric function of quantile and continuous nonparametric predictors $B_M(t,p), S_{\text{age}}(\text{age}_i,p)$, and $S_{\text{BMI}}(\text{BMI}_i,p)$. 

The functional outcomes are dominated by lower order distributional moments, resulting in monotonic quantiles in nearly all posterior samples. For the instances when the monotonic constraint is not satisfied in posterior draws, a \textit{projection-posterior} adjustment is applied to ensure monotonicity \citep{chakraborty2021coverage,wang2023coverage,lin2014bayesian}. We project the non-monotonic draw $Q^{(r)}_{Y}(p)$ onto a monotonic function based on I-splines \citep{ramsay1988monotone,wang2023splines2} using a constrained L2 loss and replace the posterior draw with its monotonic corrected function. We found that $20$ I-splines based on equally spaced knots were adequate for monotonicity adjustments.

\subsection{Bayesian Modeling Details and Posterior Inference}
\label{sec:BayesFMM}
Models \eqref{eq:quantilecoef} simplifies to a series of mixed regression models in the coefficient space. Each DR matrix originates from the respective B-spline matrix, where $J=5$ knots are selected based on quantiles of the observed continuous variable and DR matrices are equivalent across models \citep{ruppert2003semiparametric}. We fit these models using the Bayesian functional mixed models MCMC framework that has been developed in recent years with vague conjugate priors for $\tau^{\star}_{\cdot,k}$ and $B^*_{\cdot,k}$ \citep{morris2006wavelet,zhu2011robust,zhu2012robust,meyer2015bayesian,zhang2016functional,zhu2018robust,lee2018bayesian}.

Using posterior MCMC draws, denoted with superscript $(r)$, of coefficients from model \eqref{eq:quantilecoef}, we obtain $\boldsymbol{Q}^{(r)}$ conditioned on covariates. We can sample from the quantile given covariates by $\boldsymbol{Q}^{(r)} \sim f ( \boldsymbol{Q} \mid \widetilde{\boldsymbol{M}}, \widetilde{\boldsymbol{X}} )$. These posterior samples in the quantile domain can be mapped to provide posterior samples for any desired distributional summaries through different \textit{functionals}. For example, the mean of Box-Cox transformed counts $\mu_Y = \mathbb{E}[Y]$ are sampled as $\mu_Y^{(r)} = \int^1_0 Q_{Y}^{(r)}(p) dp$ where $Q_{Y}^{(r)}(p) = \boldsymbol{Q}^{(r)}$ and the original quantiles can be calculated by applying the inverse Box-Cox transform $Q_{\text{Count}}^{(r)}(p) = \text{BC}^{-1}\{ Q_{Y}^{(r)}(p) \}$. More derivations of moments are given by \cite{yang2020quantile}. Posterior samples of specific CDF values $p_\delta^{(r)}=Q^{-1 (r)}_{\text{Count}}(\delta)=\sup\{p:Q_\text{Count}^{(r)}(p)<\delta\}=\sup\{p:Q_Y^{(r)}(p)<\text{BC}(\delta)\}$ can be calculated to perform inference on the proportion of activity levels less than a specified threshold $\delta$.  For example, $\delta=1148$ corresponds to the Evenson cutpoint for moderate-to-vigorous physical activity \citep{evenson2008calibration} for a 30-second interval, and thus $1-p_{1148}^{(r)}$ comprises posterior samples for the proportion of moderate-to-vigorous physical activity, a commonly used summary in practice.  To obtain inference on the proportion of zero activity levels, we could compute $p_\text{zero}^{(r)}=\sup\{p:Q_\text{Count}^{(r)}<1\}=\sup\{p:Q_Y^{(r)}(p)<0\}$ based on our proposed Box-Cox transformation for which $Y<0$ maps to $\text{Count}<1$.

Posterior samples can be used for any desired Bayesian inference in the quantlet or quantile domain, or for any distributional summary.  In our analysis, we will focus on posterior means and credible intervals. To account for multiple testing across $p$, we will compute joint credible bands and invert them to calculate simultaneous band scores, \textit{SimBAS} \citep{meyer2015bayesian}.


\section{Simulation Study}
\label{sec:sim}

We conducted simulation studies to assess the bias induced by missingness patterns on distributional regression inference, and the effectiveness of our proposed regression-based approach in adjusting for the missingness patterns to correct that bias. We used the data outlined in Section \ref{sec:data} to simulate realistic data for our simulation study. 
We constructed a pseudo-generative model to generate realistic activity profiles by fitting these data with the Bayesian wavelet-based functional mixed model of \citep{morris2006using}
\begin{equation} \label{eq:wFMM}
\begin{array}{rl}
g \left[ \text{Count}_{i,j}(t) \right] = Y_{i,j}(t) =& \boldsymbol{X}_{i,j}^\top \boldsymbol{\eta}(t) + \mathcal{S}_\text{age}(\text{age}_{i,j},t) + \boldsymbol{Z}^\top_j \boldsymbol{\mathscr{U}}(t) + E_{i,j}(t) \\
\boldsymbol{Y}^*_{i,j} =& \boldsymbol{X}^\top_{i,j} \boldsymbol{\eta}^* + \boldsymbol{Z}_{\text{age},i,j}^\top \boldsymbol{\mathscr{U}}^*_{\text{age}} + \boldsymbol{Z}_{j}^\top \boldsymbol{\mathscr{U}}^* + \boldsymbol{E}^*_{i,j}
\end{array}
\end{equation}
using only age and sex as predictors to model $\text{Count}(t)$ on domain 6am--11:30pm. We use the transformation $g(x) = \log(x+1)$. Here, the days are indexed $i$, subjects are indexed by $j\in \{1,\dots,J\}$ and subject-specific random effects are denoted by $\boldsymbol{Z}^\top_j \boldsymbol{\mathscr{U}}(t)$ where $\boldsymbol{Z}_j$ are a one-hot vectors designed to toggle subject random intercepts. We used a smooth nonparametric effect of age, $\mathcal{S}_\text{age}(\text{age}_{i,j},t)$. We model equation \eqref{eq:wFMM} in the wavelet coefficient space denoted with superscript $^*$ using a functional decomposition analogous to the procedure described in Section \ref{sec:quantile}. The residual processes in the wavelet space are distributed as $\boldsymbol{\mathscr{U}}^* \sim \mathrm{N}(\mathbf{0},\mathbf{Q}^*)$ and $\boldsymbol{E}^*_{i,j} \sim \mathrm{N}(\mathbf{0},\mathbf{S}^*)$. Similar to posterior predictive sampling, we can simulate a population of $J=100$ subjects by first sampling 100 random effects as $\boldsymbol{\mathscr{U}}_j^* \sim \mathrm{N}(\mathbf{0},\widehat{\mathbf{Q}}^*)$. Within each subject $j$, we simulate 3 days of data by sampling $\boldsymbol{Y}^*_{i,j} \sim \mathrm{N}( \widehat{\boldsymbol{\mu}}_{j} + \boldsymbol{\mathscr{U}}_j^*, \widehat{\mathbf{S}}^*)$ 3 times where $\widehat{\boldsymbol{\mu}}_{j} = \boldsymbol{X}^\top_{j} \widehat{\boldsymbol{\eta}}^* + \boldsymbol{Z}_{\text{age},j}^\top \widehat{\boldsymbol{\mathscr{U}}}^*_{\text{age}}$. Note that the hat script denotes posterior mean estimates of \eqref{eq:wFMM}. The vector $\boldsymbol{X}^\top_{j}$ is modified to reflect a different combination of age and sex for each subject $j$. For each replicate, we simulate 100 subjects for each sex by age combination for ages 12--16, resulting in $N=1000$ total subjects with 3 days each. Each simulation in the wavelet space, $\boldsymbol{Y}^*_{i,j}$ was then transformed back to the time domain $\text{Count}_{i,j}(t)$ and rounded to the nearest non-negative integer. Our pseudo-generative model draws of $\text{Count}_{i,j}(t)$ are presented in supplemental plots and show reasonably realistic-looking activity profiles.  Our simulation methodology is distinct from the quantile functional regression outlined earlier, so no quantlet space assumptions were made in the generation of the data. We freely share our simulation code and simulated data in online supplements for others to use in their projects.

We generated complete data with no missingness, and then induced missingness according to two different missing data patterns. We now outline our procedures for generating the missing data labels for the simulation. 

\subsection{Case 1: Bed-Time Missing Data}

We denote early morning and late night missing data as the \textit{bed-time} missing pattern. For early mornings, we used the empirical data from Figure \ref{fig:heat} to estimate a truncated normal distribution of the length of missing data for each age group. For each day within an age group, we sample a length of missing data from the truncated normal and zero out those observations. We repeat the analogous sampling procedure the late night missing data and then apply the \textit{30 minute rule} to obtain missing data labels. We present a heatmap of the training dataset in the supplement for the bed-time simulation. Based on Case 1 training data, we trained a Box-Cox $\lambda = -0.02020202$ and elected to use a sparse $K_Q=7$ quantlet system and $K_M=11$ functional PCs.

\subsection{Case 2: Bed-Time and Day-Time Missing Data}

In addition to the missing data patterns of Case 1, we also apply variable day-time missing patterns based on sex. Given a day specific missing indicator variable, the missing data pattern is applied to the accelerometer data for that day. Using a normal distribution centered a 6pm with a standard deviation of 2 hours, we sample a missing data run length of missing data to be replaced by zeros. The missing indicator variable for each day is sampled from a $\mathrm{Ber}(0.3)$ and $\mathrm{Ber}(0.2)$ for males and females, respectively.  We then apply the \textit{30 minute rule} to obtain missing data labels and a heatmap of the training dataset is presented in the supplement for the day-time simulation. Based on Case 2 training data, we trained a Box-Cox $\lambda = -0.02020202$ and elected to use a sparse $K_Q=7$ quantlet system and $K_M=13$ functional PCs. 

\subsection{Simulation Results}
We have the simulated \textit{true} data without the applied missing data labels to use as the benchmark, the posterior predictive samples of equation \eqref{eq:wFMM}. We calculate the Frechet mean, $\mathbb{E} \left[ Q_Y(p) \mid \mathrm{age},\mathrm{sex}, \pi_0(t)\right]$, for each age and sex group across 500 replicates without missing data labels as the truth. Within each replicate data set, we have 100 subjects for each age-sex combination and 50000 total subjects to accurately estimate $\mathbb{E} \left[ Q_Y(p) \mid \mathrm{age},\mathrm{sex}, \pi_0(t)\right]$. We fit our quantlet functional regression model and assess the recovery of the true Frechet means and summary statistics using $\pi_0(t)=0$ as the missing rate predictor.

To assess the benefit of our unified quantile functional modeling approach borrowing strength across $p$ using a functional regression approach, we compare with a competing method that models a subset of quantiles independently with no borrowing of strength across $p$. We fit 13 independent functional linear array models (FLAMs) at $p \in \left\{ 1/1025, 0.01, 0.025, 0.05, 0.10, 0.25, 0.50, 0.75, 0.90, 0.95, 0.975, 0.99, 1024/1025 \right\}$ as independent linear mixed effect models using the same nonparametric smooth effects and functional predictors outlined in Section \ref{sec:quantile} \citep{brockhaus2015functional}. The estimated conditional expectations of quantile points for the FLAM, at each age and sex group, were interpolated to provide an estimate of the Frechet mean \citep{borchers2023pracma}. 

For evaluation, we apply the following competing models:
\begin{enumerate}
    \item The quantlet functional regression (QFR) model \eqref{eq:quantilecoef} \textit{with missing adjustment}: including the missing data profile predictor $\pi(t)$ as a model parameter 
    \item QFR \textit{without missing adjustment}: not including missing data profile predictor $\pi(t)$
    \item Independent FLAM \textit{with missing adjustment:} including $\pi(t)$
    \item FLAM \textit{without missing adjustment:} not including $\pi(t)$ 
\end{enumerate}

\begin{figure}[H]
\centering
\includegraphics[width=6.25in]{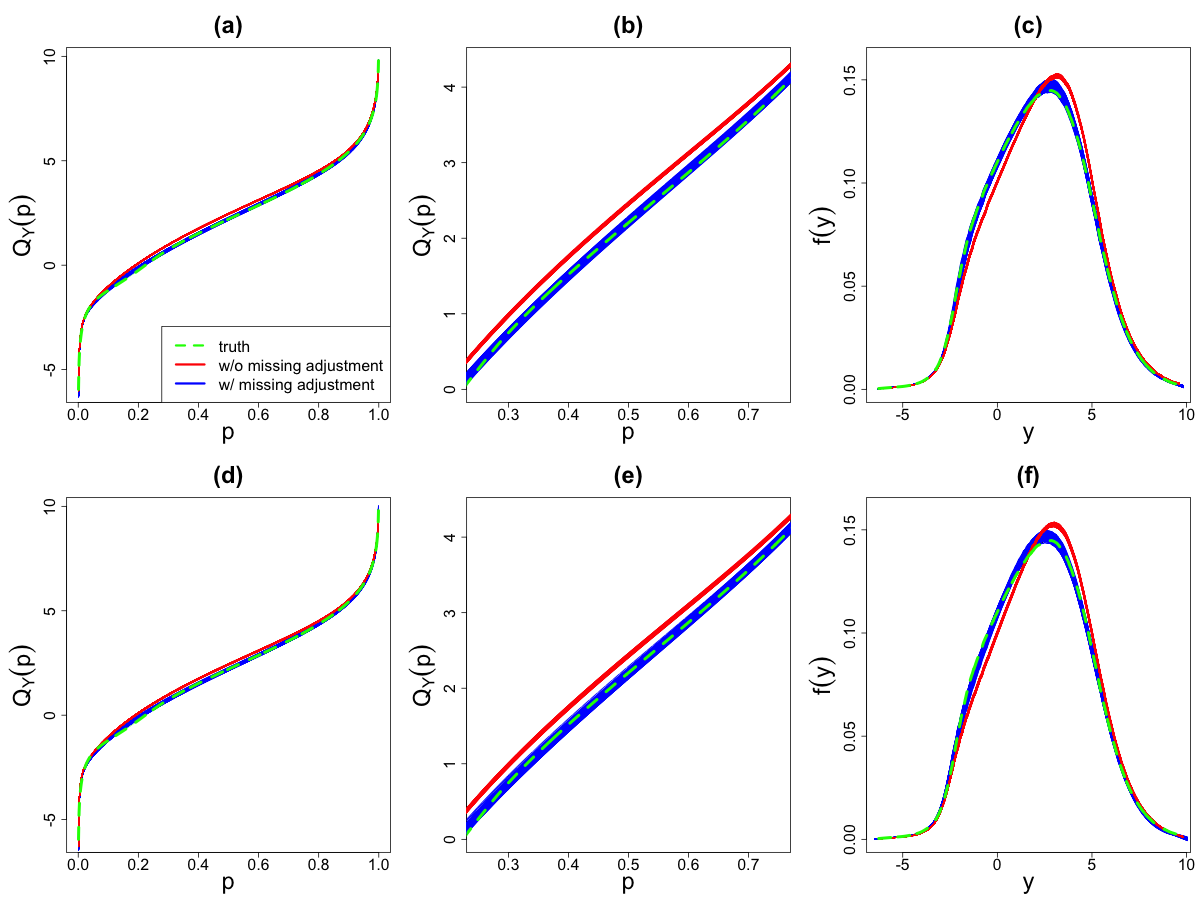}
\caption{Quantile function estimates for $\text{age}=14$ males from 500 replicates. Model 1 estimates are plotted in blue, Model 2 estimates are plotted in red and true quantile are plotted in green. (a) Case 1 quantile function estimates. (b) Case 1 quantile function estimates on domain $p \in [0.25,75]$. (c) Case 1 density based on transforming quantile function estimates. The same set of plots for Case 2 are presented in (d)--(f).}
\label{fig:sim_results}
\end{figure}

\begin{figure}[H]
\centering
\includegraphics[width=6.25in]{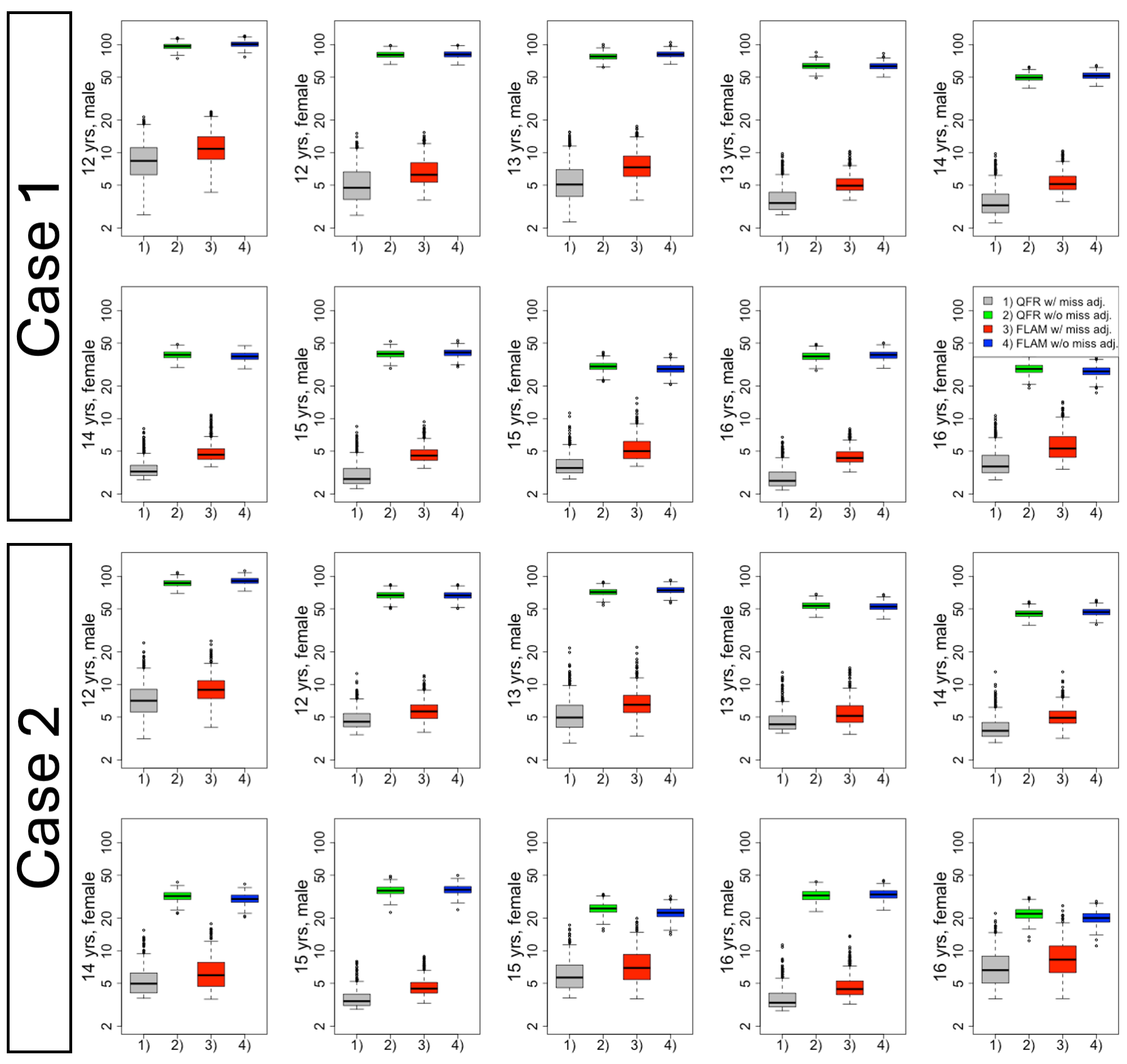}
\caption{Integrated square errors (ISEs) on the log scale for age and sex combinations under competing methods. 1) Quantile functional regression (QFR) with missing data adjustment. 2) QRF without missing data adjustment. 3) Functional linear array model (FLAM) with missing data adjustment. 4) FLAM without missing data adjustment.}
\label{fig:ise}
\end{figure}

For each age and sex group, we compared the integrated square error (ISE) given as $\begin{array}{c}
\mathrm{ISE}(\boldsymbol{Q}) =
{ \left\| \widehat{\boldsymbol{Q}}_r - \mathbb{E} \left[ \boldsymbol{Q} \right] \right\|_2^2 }
\end{array}$ and abbreviate $\mathbb{E} \left[ Q_Y(p) \mid \mathrm{age},\mathrm{sex},\pi_0(t)\right] = \mathbb{E} \left[ \boldsymbol{Q} \right]$. Example estimates of $\widehat{\boldsymbol{Q}}_r$ are plotted in Figure \ref{fig:sim_results} and ISE results are plotted in Figure \ref{fig:ise}.

First, comparing Models 1 and 3 to Models 2 and 4, we see the benefit of the basis space modeling for improved estimation accuracy. The QFR modeling the entire quantile function using quantlets had lower ISE than the FLAM performing independent quantile-by-quantile analysis, for models that took into account missing data (Models 1 vs. 3) or ignoring missing data (Models 2 vs. 4), in spite of the fact that the QFR only modeled 7 quantlets while the FLAM was fit independently to 13 different quantiles.  We believe this improvement demonstrates the benefit of using a unified functional data analysis approach that explicitly borrows strength across $p$ according to the learned basis functions.

Second, when comparing Models 1 \& 3 vs. 2 \& 4, we see the great improvement obtained when accounting for the missing data using our regression/normalization strategy.  For both the QFR and the FLAM, the inclusion of the missing data profile as a covariate greatly improved the estimation of the true mean quantile functions for each age-sex group.  We believe that this result reflects the ability of our proposed approach to adjust for potential bias caused by the missingness, with the model learning from the observed data which times of day tend to have higher activity levels and accounting for this tendency by adjusting the predicted distributions based on these patterns to mitigate this bias.
Figure \ref{fig:sim_results} shows that accounting for the missing rate aligns the estimated quantiles with their true values, and failure to account for the missing rate skews the distribution to the right. Notably, adjusting for missing data patterns allow the nonparametric age effect and functional coefficients for sex to be more accurately estimated.

\begin{figure}[H]
\centering
\includegraphics[width=6.25in]{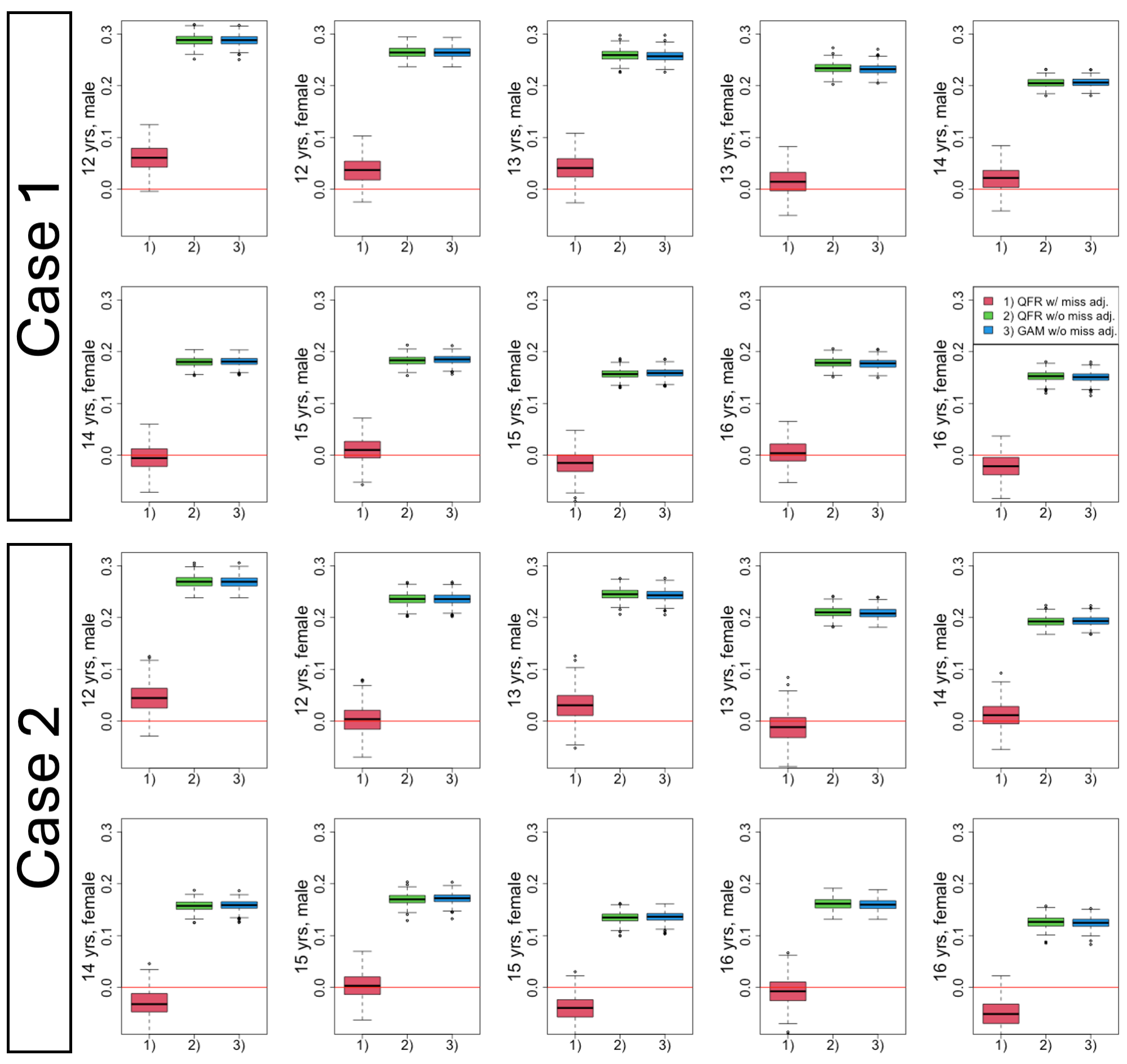}
\caption{Bias of estimating $\mu_Y$ for age and sex combinations under competing methods. 1) Quantile functional regression (QFR) with missing data adjustment. 2) QRF without missing data adjustment. 3) Generalized additive model (GAM) without missing data adjustment.}
\label{fig:first_moment}
\end{figure}

Given that the missing data bias is also expected to affect distributional summaries, we compared our QFR with and without the missingness adjustment with a feature extraction approach that computes the mean activity level for each individual and relating to covariates using a GAM with the same predictors as our QFR \citep{wood2017GAM}.  We computed estimates of $\mu_Y=\int_0^1 Q_Y(p)$ for each age-sex group in the QFR model, with and without missing data adjustment, and the age-sex specific mean estimates from the GAM and compared with the true means from the pseudo-generative distribution with no missing data, with the deviations plotted in Figure \ref{fig:first_moment}.  First, note that the QFR has similar performance to the GAM when missing data is ignored, demonstrating that there is no substantial loss of efficiency with our unified approach performing functional regression on the entire distribution and then looking at means as a post-hoc analysis versus a feature extraction approach directly modeling the mean.  More importantly, if the missingness patterns in the data are ignored we see observe a high level of bias in both QFR and GAM-based mean analyses, but this bias is notably mitigated by our regression/normalization based missing data strategy.  This demonstrates that our distributional modeling approach with missing data adjustment can obtain better inference on distributional summaries than standard approaches directly modeling these summaries that do not account for the nuanced missing data patterns in these activity data.

\section{Analysis of TEAN Study Accelerometer Data}
\label{sec:real}

We assess statistical significance of covariate effects 
based on the simultaneous band scores (SimBaS) 
computed by inverting the joint credible bands using the procedure outlined in \cite{meyer2015bayesian}. We discuss its implementation in detail in the supplement. 
The SimBaS we construct, $\boldsymbol{P}_{\operatorname{SimBaS}}$, 
are local probability scores that account for multiple testing across $p$ when determining which covariates significantly affect the activity distribution and mark which regions of $p$ significantly differ.

To explore the effect of missing data, we estimate predicted quantile functions with different missing data patterns and other covariates at their cell mean values $\widetilde{\boldsymbol{X}} = (0.25, 0.25, 0.25,\\ 0.25, 14, 50)$, averaging over sex and site as well as assuming $\text{age}=14$ and median BMI.  We illustrate the marginal effect of how missing data at 1.5 hour increments at different time points affects the composition of different activity levels in the data. 

\begin{figure}[H]
\centering
\includegraphics[width=4in]{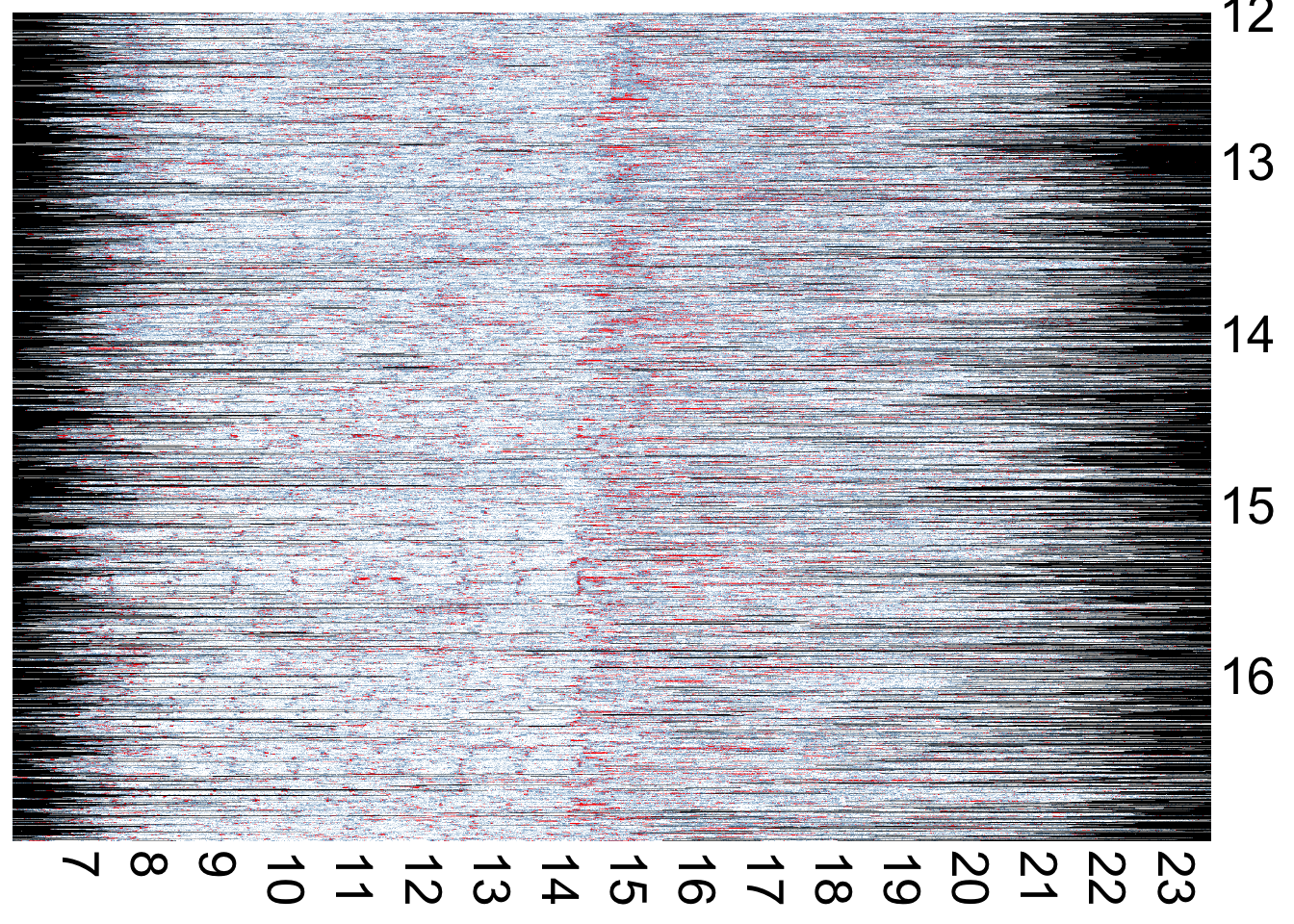}
\caption{\small Heatmap of accelerometer count at 30s epochs from plotted on a x-axis of 6am--11:30pm. TEAN cohort subjects ($N=750$) were order by age, with 12yr (18\%), 13yr (21\%), 14yr (20\%), 15yr (21\%), 16yr (20\%), 17yr ($<1$\%). Consecutive epochs $\geq 30$ minutes of zero counts are labeled missing. Epochs of missing data are colored black, accelerometer counts are colored in shades of blue and accelerometer counts above the moderate-vigorous cutoff are colored red.}
\label{fig:heat}
\end{figure}

\begin{figure}[H]
\centering
\includegraphics[width=5.6in]{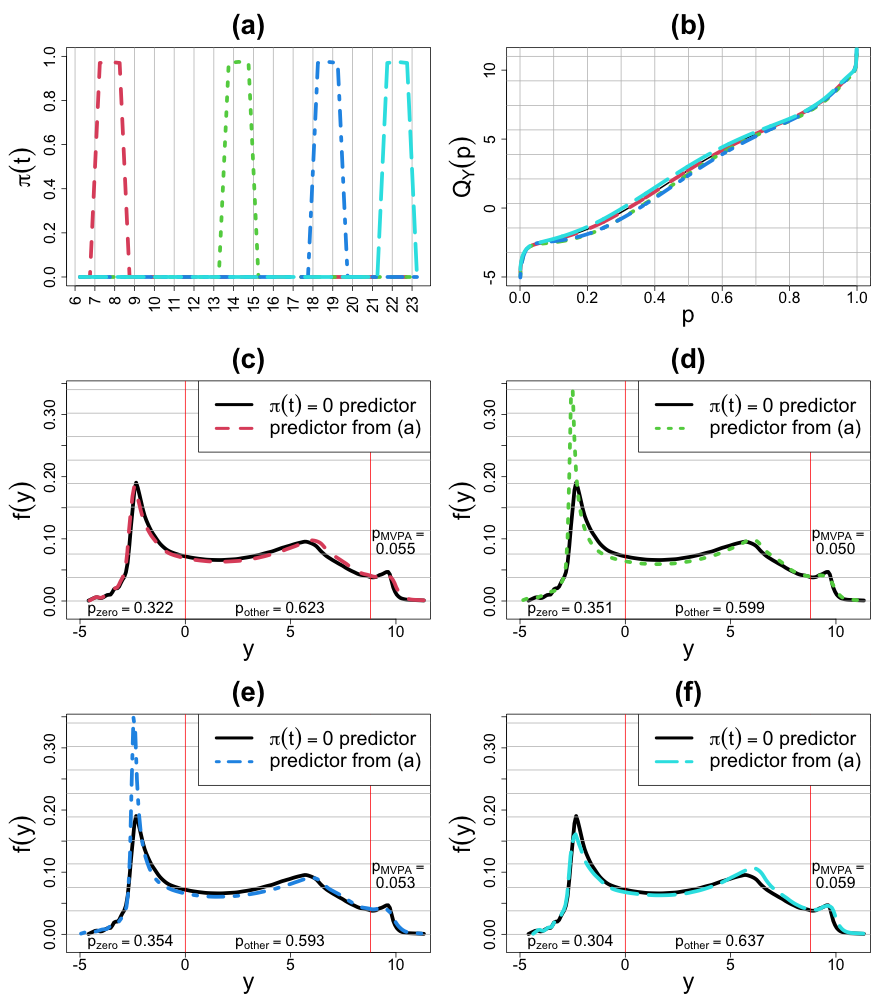}
\caption{Marginal missing data effects. (a) Plot of 4 missing data predictors. (b) Estimated quantile function in Box-Cox transformed space based on 4 missing data predictors, matched by line style. (c)--(f) Quantile functions transformed back to densities with demarcations for zero counts and moderate-to-vigorous physical activity (MVPA) cutoffs. For reference, contrasting densities estimated using zero missing data ($\pi(t)=0$) are plotted in black.}
\label{fig:miss_effect}
\end{figure}

Figure \ref{fig:miss_effect} presents predicted quantile functions for various missing data patterns with the corresponding density functions, predicted proportion of zero activity, and predicted proportions of moderate-to-vigorous physical activity (MVPA) presented to enhance interpretability.
Note from the heatmap of the full data set (Figure \ref{fig:heat}) that MVPA values concentrate around 2-3pm and missing data around 2-3pm decreases the overall proportion of MVPA (Figure \ref{fig:miss_effect}d). In addition, activity levels below MVPA are observed at 10pm, and missing data at 10pm increase the proportion of MVPA (Figure \ref{fig:miss_effect}f). Also note that data at 10pm is predominantly zero counts and a decrease in the proportion of zeros is observed when data is missing at this time. These changes in the composition of the activity data can lead to bias when computing summary statistics and, as demonstrated by our simulation studies, are remedied by our strategy of covariate adjustment for missing data patterns. Additionally, note that subtle changes in the proportions of physical activity due to missing data are amplified when these proportions are used to calculate total minutes of MVPA, a commonly-used summary for activity data in the literature. Dynamic visual representations of the impact of different missing data patterns outlined in Figure \ref{fig:miss_effect}a on the distribution are available in the Supplemental Materials. The missing data adjustment based on a smooth function-on-function regression results in seamless transitions of quantile estimates between similar missing data patterns.

\begin{figure}[H]
\centering
\includegraphics[width=5.5in]{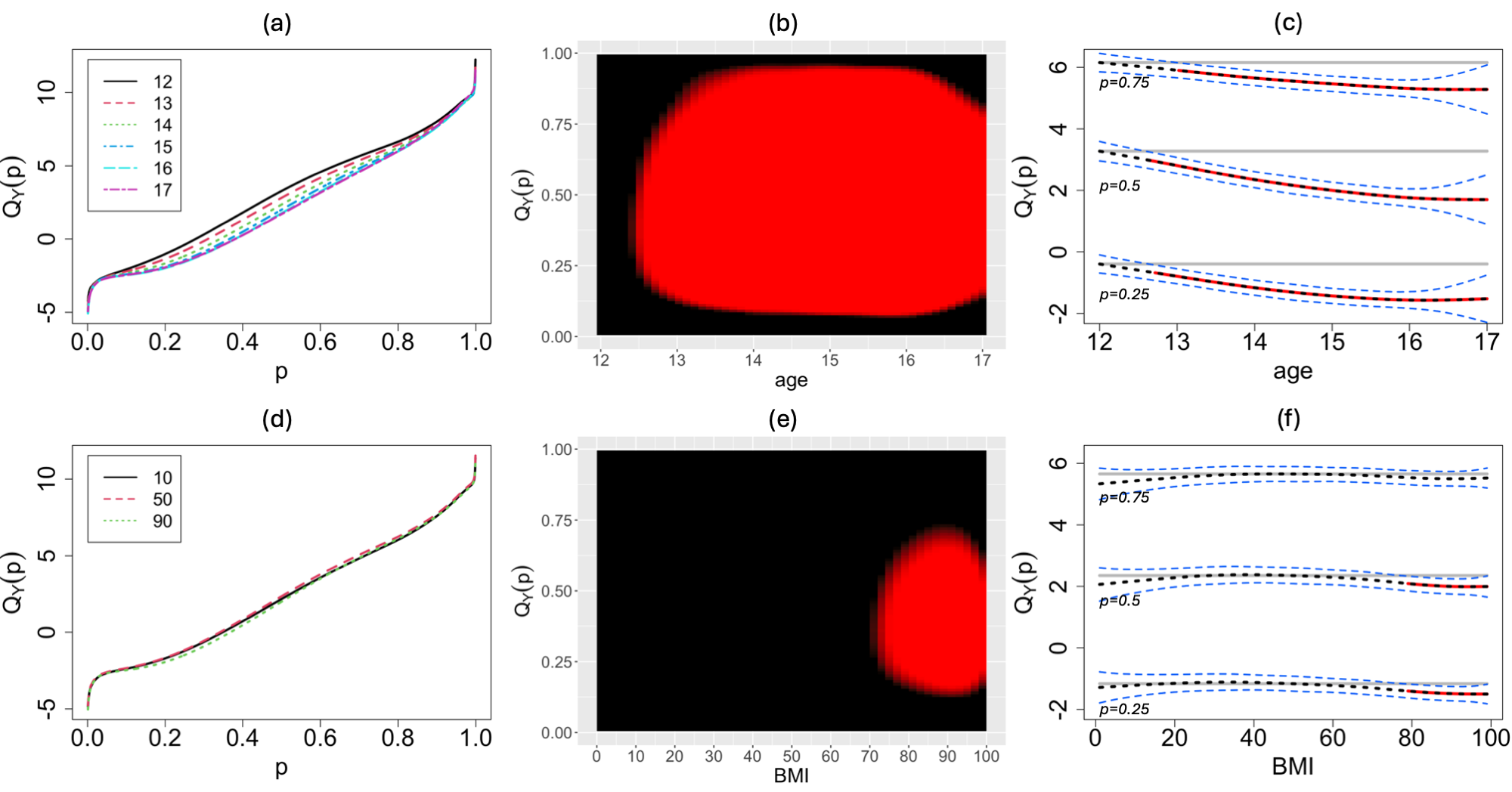}
\caption{Quantile function estimates based on smooth nonparametric effects of age and BMI. (a) Posterior quantile estimates for 6 age groups. (b) Age-by-$p$ heatmap of $\boldsymbol{P}_{\operatorname{SimBaS}}$ to test $\boldsymbol{Q}(\text{age}) = {\boldsymbol{Q}}( \text{age}=12 )$ with red denoting the significant $\boldsymbol{P}_{\operatorname{SimBaS}}<0.05$. (c) Cross section of posterior quantile functions as a function of age at $p \in \{0.25,0.50,0.75\}$. Dotted lines denote the posterior means and joint bands, horizontal gray lines denote the null hypothesis  $\boldsymbol{Q}(\text{age}) = {\boldsymbol{Q}}( \text{age}=12 )$ and red denotes regions with significant $\boldsymbol{P}_{\operatorname{SimBaS}}$. (d)--(f) are the analogous plots for nonparametric BMI \textit{percentile} effects with hypothesis test $\boldsymbol{Q}(\text{BMI}) = {\boldsymbol{Q}}( \text{BMI}=50 )$.}
\label{fig:jmaps}
\end{figure}

Assured that our missing data adjustment mitigated distributional bias in the data, we now turn to an assessment of how activity levels systematically vary by 
age and BMI (Figure \ref{fig:jmaps}a). To represent 
age effects under an average missingness pattern, we fixed $\widetilde{\boldsymbol{M}} = \mathbf{0}$ (the sample mean missing data pattern), $\widetilde{\boldsymbol{X}} = (0.25, 0.25, 0.25, 0.25, \text{age}, 50)$ with $\mathrm{BMI}=50$ (the population median) and let age vary. We denote these quantile estimates that vary with age as $\boldsymbol{Q}(\text{age})$.  We use the simultaneous test procedure outlined in the supplement to assess distributional differences between $\boldsymbol{Q}(\text{age})$ for $\text{age} \in [12,17]$ and ${\boldsymbol{Q}}( \text{age}=12 )$ derived from the posterior mean estimate. Figure \ref{fig:jmaps}b shows the region flagged with $\boldsymbol{P}_{\operatorname{SimBaS}}<0.05$, at what age the quantile function is significantly different from the ${\boldsymbol{Q}}( \text{age}=12 )$ quantile function. Figure \ref{fig:jmaps}c compares the smooth quantile function by age effects at $p \in \{0.25,0.50,0.75\}$ with age=$12$. In general, as age increases, we observe a decrease in accelerometer activity at the 25th and 50th quantile. We observe significant decreases in right tail near the 75th quantile for 16- and 17-year olds. Though we look at particular values of $p$, this probability scoring procedure can be applied to any value of $p$ and distributional summary statistics.

We computed a similar analysis on BMI, conditioning on $\text{age}=14$ 
(see Figure \ref{fig:jmaps}d--f). We compared the activity level distributions for the various BMI with $\boldsymbol{Q}( \text{BMI}=50 )$, the 50th percentile of adolescent BMI. High BMI ($\mathrm{BMI} \approx 100$) is associated with decreased activity at the 25th and 50th quantiles, but does not change activity at the 75th quantile. The results for the 25th quantiles can be interpreted as an increase in the proportion of sedentary activity occurs for high BMI percentiles. Low BMI ($\mathrm{BMI} \approx 0$) is also associated with reduced physical activity but has large standard errors due to few $\mathrm{BMI} < 20$ observations. Results for estimates based on sex and site can be found in the supplement. Code and data to reproduce the TEAN analysis are provided in the supplemental material.

Note that the left and right tail extreme values of the empirical distribution (Figure \ref{fig:BC} \& \ref{fig:jmaps}a) are consistent across the population. Forcing a linear effect for age, $\mathrm{age} \times \beta_{\mathrm{age}}(p)$, would significantly attenuate the tails of the quantile by a multiplicative factor of age and provide biased inference based on the misspecified linear assumptions. We observe wider joint bands at $\mathrm{age} \approx 17$, $\mathrm{BMI} < 20$ and $\mathrm{BMI} \approx 100$ due to these characteristics being sparsely represented in the dataset (Figure \ref{fig:heat}). This inference is also observed in nonparametric models and GAMs for sparsely represented continuous covariates. In a linear model, the standard errors of effects are homogeneous across age groups and fail to take into account sparsely represented age groups. A smooth nonparametric effect is a more flexible alternative in many distributional modeling settings and an important feature of our proposed modeling framework.

\section{Discussion}

We propose novel approaches for distributional regression of subject-specific activity level distributions on covariates, allowing arbitrary numbers of discrete and continuous covariates, allowing smooth and not just linear covariate effects on the distribution, and enabling inference at the distributional level as well as post-hoc analyses of any desired distributional summary statistics. We used a functional regression approach with sparse, near-lossless quantlet basis representations, which as demonstrated by simulation yields more efficient estimation and inference than independent models for a grid of quantiles. We also introduced the use of covariate adjustment of missing data patterns via functional predictors to study how missing data affects the shape of quantile functions, and to adjust for this bias and effectively calibrate the data across the observed subject-specific missing data patterns in our distributional regression. Our simulation studies showed that this approach greatly reduced the bias induced by the missing data which affects existing approaches involving distributional modeling or distributional summary statistics that ignore the missing data. This reduction in bias is of critical importance for data from accelerometers and other wearable devices, as missing data is commonly encountered and this missingness can bias the empirical distributional estimates as well as any of their summaries estimates. 

The regression/normalization strategy we introduce for adjusting for missing data patterns can be adapted to other distributional data analysis settings, and the modeling approach we introduce here can be applied to other wearable device settings {or other settings yielding high-frequency data streams}. The strategy of using missingness patterns as functional covariates to calibrate empirical subject-specific distributions introduced here is promising, and should be further studied in {this and} other settings.

\section{Data Availability}
TEAN study data and supplemental materials are available upon request. 

\begin{figure}[H]
\centering
\singlespacing
\includegraphics[width=4.5in]{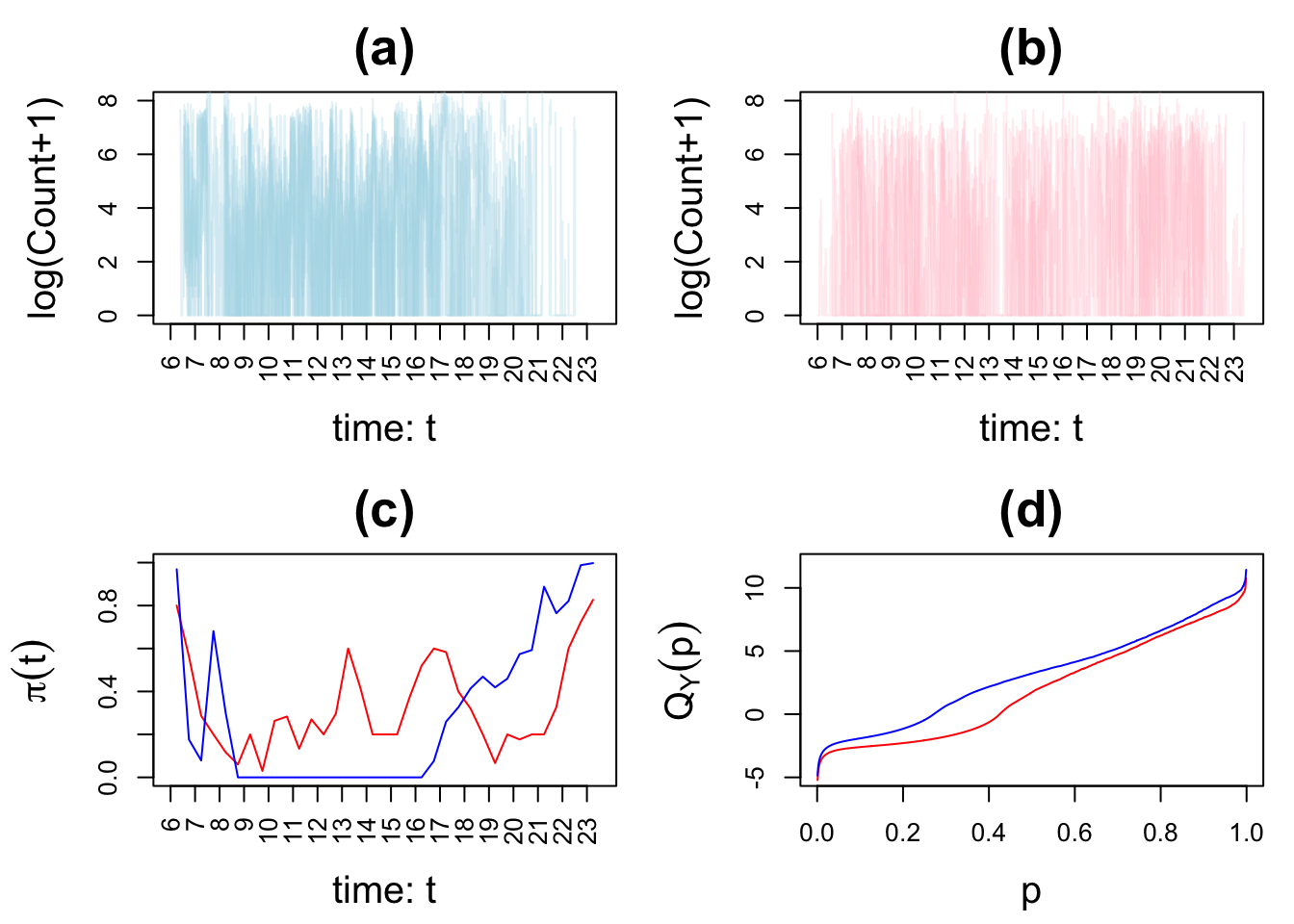}
\caption{Examples of accelerometer data from the study cohort. (a) Subject with missing data during the mornings and evenings. (b) Subject with missing data during the middle of the day. (c) Subjects' missing data rates by time-of-day aggregated at half-hour bins. (d) Quantile functions of transformed count distributions.}
\label{fig:miss}
\end{figure}

\bibliographystyle{agsm}
\bibliography{Bibliography}

\end{document}